# Cysteine and cystine adsorption on FeS$_2$(100)

Teppei Suzuki [a,*], Taka-aki Yano [b], Masahiko Hara [b], and Toshikazu Ebisuzaki [a]

[a] *Computational Astrophysics Laboratory, RIKEN, 2-1 Hirosawa, Wako, Saitama 351-0198, Japan*
[b] *Chemical Evolution Lab Unit, Earth-Life Science Institute (ELSI) and Department of Chemical Science and Engineering, School of Materials and Chemical Technology, Tokyo Institute of Technology, 4259-G1-7 Nagatsuta, Midori-ku, Yokohama 226-8502, Japan*

[*] Corresponding author
*E-mail address:* teppei.suzuki@riken.jp

(December 18, 2017)

ABSTRACT

Iron pyrite (FeS$_2$) is the most abundant metal sulfide on Earth. Owing to its reactivity and catalytic activity, pyrite has been studied in various research fields such as surface science, geochemistry, and prebiotic chemistry. Importantly, native iron–sulfur clusters are typically coordinated by cysteinyl ligands of iron–sulfur proteins. In the present paper, we study the adsorption of L-cysteine and its oxidized dimer, L-cystine, on the FeS$_2$ surface, using electronic structure calculations based density functional theory and Raman spectroscopy measurements. Our calculations suggest that sulfur-deficient surfaces play an important role in the adsorption of cysteine and cystine. In the thiol headgroup adsorption on the sulfur-vacancy site, dissociative adsorption is found to be energetically favorable compared with molecular adsorption. In addition, the calculations indicate that, in the cystine adsorption on the defective surface under vacuum conditions, the formation of the S–Fe bond is energetically favorable compared with molecular adsorption. Raman spectroscopic measurements suggest the formation of cystine molecules through the S–S bond on the pyrite surface in aqueous solution. Our results might have implications for chemical evolution at mineral surfaces on the early Earth and the origin of iron–sulfur proteins, which are believed to be one of the most ancient families of proteins.

*Keywords:* Pyrite; Cysteine; Cystine; Adsorption; Prebiotic chemistry; Origin of life

(Submitted to *Surface Science*)

## 1. Introduction

Iron pyrite (FeS$_2$), or fool's gold, is the most abundant metal sulfide on Earth [1] and has long been studied as an important material in geochemical and environmental processes [2–5]. In materials science, because of its high abundance and low material costs, together with recent progress in nanotechnology [6], pyrite is a promising material for chemical energy storage [7–11] and solar energy conversion [12, 13]. Furthermore, pyrite has been studied in the research field of the origins of life [14–19]; owing to its high reactivity and catalytic activity [1], the material could have been a concentrator of prebiotic molecules and a catalyst for polymerization in the early evolution of life. The idea, first proposed by Wächtershäuser [20–23], is known as the 'iron–sulfur world' hypothesis; it proposes that mineral surfaces could have facilitated chemical reactions under anoxic conditions in the early history on Earth. Hence, understanding the reactivity and adsorption of organic molecules on minerals surfaces (such as pyrite surface) is of fundamental importance in surface science and prebiotic chemistry.

In addition, iron–sulfur proteins that most frequently contain F$_2$S$_2$, Fe$_3$S$_4$, and Fe$_4$S$_4$ clusters undergo oxidation-reduction reactions in biologically important processes such as photosynthesis, nitrogen fixation, and electron-transfer chains in respiration [24–26]; Fe–S clusters act as catalytic centers in the enzymes that promote these reactions. Native iron–sulfur clusters are typically coordinated by cysteinyl ligands of iron–sulfur proteins such as ferredoxin [24–26]. With this in mind, we draw attention to the interaction of iron sulfide with cysteine in the context of the origin of enzymatic proteins. In fact, Fe–S clusters are common to the most ancient and primitive living organisms and thus ferredoxins are believed to be one of the most ancient families of proteins [24–26].

Experimental and theoretical studies have shown that amino acids are strongly absorbed on surfaces of noble metals (gold [27–33], silver [34–38], and cupper [39–41]), metal oxides [42–44], and metal sulfide [45–47]. In recent years, experiments under ultra-high vacuum (UHV) conditions have been applied to the investigation of the adsorption onto pyrite surfaces [45–47]; these studies have provided some insights into the adsorption process on pyrite surfaces. Despite many efforts, however, our understanding on how amino acids are adsorbed at the atomic level remains limited. Amino acids contain the carboxyl (COOH) and the amino ($NH_2$) groups; they also contain the headgroup (except for glycine). The pyrite surface, on the other hand, has two possible interacting sites: the Fe and S atoms. This unique nature of pyrite surface coupled with the complexity of the molecular structures of amino acids could make the adsorption properties highly non-trivial. Computational approaches based on electronic structure calculation methods, such as density functional theory (DFT) [48, 49], should in principle shed light on the adsorption mechanisms. Unfortunately, only a few computational studies on the adsorption of glycine on pyrite have been reported [50, 51].

In the present study, using DFT calculations and Raman spectroscopy measurements, we investigate the adsorption of L-cysteine and its oxidized dimer, L-cystine, on the $FeS_2$(100) surface, which is the most stable pyrite plane [52]. Cysteine is an important amino acid containing the thiol (SH) headgroup; it is involved in many biological processes including the folding and stability of proteins [53, 54], redox chemistry [55], and electron transfer [24–26, 56–58]. In addition, cysteine plays an important role in accelerating the oxidation rate of pyrite [59]. To our knowledge, the present work is the first density-functional study on the interaction of iron sulfide (such as pyrite surfaces) with cysteine and cystine. The rest of the paper is organized as follows. In Methodology, computational details as well as experimental method are described. We discuss the optimized geometries of a variety of adsorption modes of cysteine (and cystine) on the ideal and/or sulfur-deficient surfaces and calculate their adsorption energies. We also discuss the results of Raman spectroscopy measurements on the cysteine/cystine adsorption on the pyrite surface in aqueous solution. Then we summarize our conclusions.

## 2. Methodology

### 2. 1. Computational details

All the electronic structure calculations were carried out based on the Kohn–Sham formation of DFT [48, 49]. The BLYP exchange-correlation functional [60, 61] was chosen and the valence–core interactions were described by Troullier–Martins norm-conserving pseudopotentials [62]. Nonlinear core correction [63] was included for the pseudopotential of the Fe atom in order to enhance transferability. In the present work, the van der Waals interaction was not included. Periodic boundary conditions were employed throughout the calculations. Previous theoretical studies have shown that the electronic state of Fe atoms in pyrite is low-spin state for bulk, ideal surface, as well as sulfur-vacancy surface [51, 64, 65], supporting the use of standard DFT methods for studying pyrite surfaces. A cutoff for plane-wave expansion was 80 Ry. Only the $\Gamma$-point was used to sample the Brillouin zone (although k-point calculations may provide more precise results, we note that $\Gamma$-point-only calculations within a supercell approach have shown to provide reasonably good results for describing pyrite surface [64].) The preconditioned conjugate gradient method was used for the optimization of wavefunctions, in which a convergence criteria for orbitals was set to $1\times10^{-6}$ hartree. All the geometries have been optimized using a maximum force tolerance of $1\times10^{-3}$ hartree/bohr ($\approx$0.05 eV/ Å).

Our slab models for iron pyrite surfaces were similar to those previously reported by other groups [51, 64, 65]. The $FeS_2$(100) surface was chosen because it is the most stable surface [52]. In modeling the cysteine–surface system, we used a supercell of $10.8562 \times 10.8562 \times 22.5$ Å$^3$ (experimental lattice parameters [66] were used for determining the cell size for $x$ and $y$ directions); and a slab model of a (2 × 2) $FeS_2$(100) surface consisting of 15 atomic layers (40 Fe and 80 S atoms) (see Fig. 1(a)). One sulfur atom (as indicated by $S_V$ in Fig. 1(b)) was removed from the clean surface to generate a sulfur vacancy on the pyrite surface. In geometry optimization, the nine bottom-most atomic layers were fixed whereas the six topmost atomic layers were relaxed. In the present work, we studied neutral or zwitterion species of a cysteine molecule. For our cystine–surface model, we used a supercell of $16.1106 \times 16.1106 \times 18.0$ Å$^3$ and a slab model of a (3 × 3) $FeS_2$(100) surface consisting of nine atomic layers (54 Fe and 108 S atoms) (see Fig. 1(c)). Two sulfur atoms (as indicated by $S_{V1}$ and $S_{V2}$ in Fig. 1(d)) were removed from the ideal surface to create the defective pyrite surface. In geometry optimization, the three bottom-most atomic layers were fixed whereas the six topmost atomic layers were relaxed. In the present work, we studied the neutral form of a cystine molecule. We used CPMD code version 4.1 [67, 68] in all the calculations. The three-dimensional visualization of the optimized geometries in this article was performed using VESTA software [69].

The adsorption energy $\Delta E_{\text{ads}}$ can be obtained by

$$\Delta E_{\text{ads}} = E_{\text{system}} - (E_{\text{slab}} + E_{\text{Cys}}), \tag{1}$$

where $E_{\text{system}}$ is the total energy of the surface with a cysteine molecule; $E_{\text{slab}}$ is the total energy of the pyrite surface (slab); and $E_{\text{Cys}}$ is the total energy of an isolated molecule that is obtained using the same supercell as used in the slab system. A similar definition is applied for cystine adsorption. In the above definition, the negative adsorption energy means that adsorption is energetically possible.

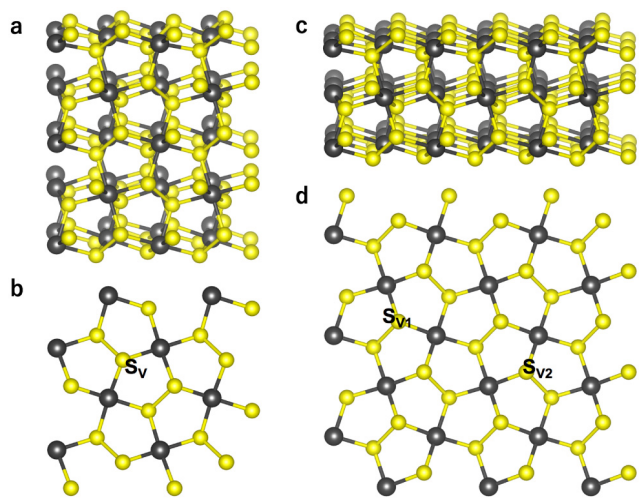

**Fig. 1.** Slab models for the $FeS_2(100)$ surface. Sulfur and iron atoms are represented by yellow and dark grey, respectively. (a) A slab model of a $(2 \times 2)$ $FeS_2(100)$ surface with 15 atomic layers (40 Fe and 80 S atoms); (b) top view of the slab (only the three topmost atomic layers are shown for clarity), where $S_V$ indicates one sulfur atom removed from the clean surface to generate a sulfur vacancy; (c) a slab model of a $(3 \times 3)$ $FeS_2(100)$ surface with nine atomic layers (54 Fe and 108 S atoms); (d) top view of the surface (only the three topmost atomic layers are shown for clarity), where $S_{V1}$ and $S_{V2}$ indicate two sulfur atoms removed from the ideal surface to create the defective surface.

*2. 2. Raman spectroscopy measurements*

The Raman spectrum was on the pyrite surface after a natural cubic crystal of pyrite was dipped into a 0.1 mM aqueous solution of cysteine monomers for 10 hours. Raman spectra were measured with the use of a line-scan confocal Raman microscope (Raman 11, Nanophoton Corp., Japan) with an excitation wavelength of 532 nm. The spatial resolution of the microscope was ~350 nm with a 0.9 NA objective lens. The laser power density and acquisition time were set to ~0.1 W/cm2 and 10 sec, respectively.

## 3. Results and discussion

Table 1
Adsorption energies and structural parameters of cysteine adsorption on the clean $FeS_2(100)$ surface. $S_{Cys}$ denotes the sulfur atom of the cysteine molecule and $O_C$ the oxygen atom of the carboxyl group.

| Configuration | $\Delta E_{ads}$ (kcal/mol) | $S_{Cys}$–Fe (Å) | $O_C$–Fe (Å) |
|---|---|---|---|
| H1 | −7.0 | 2.40 | |
| HC1 | −7.5 | 2.67 | 2.02 |
| HC2 | −10.5 | 2.51 | 2.03 |
| Z | −10.6 | | 2.13, 2.18 |

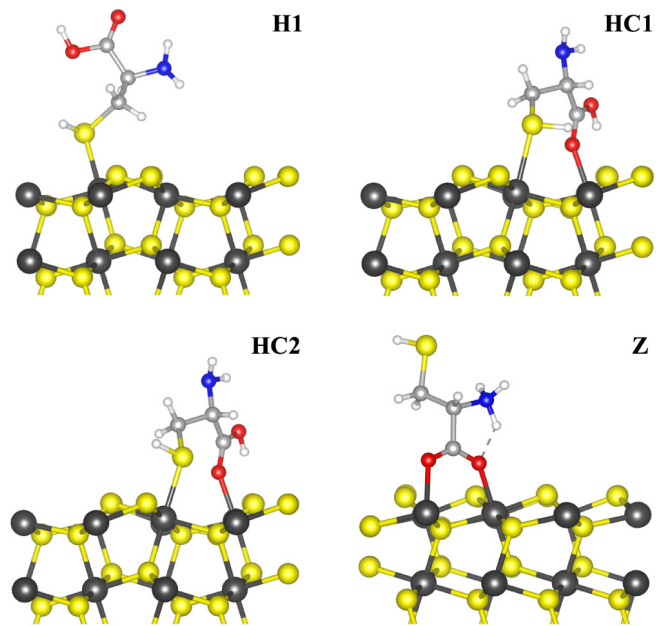

**Fig. 2.** Optimized geometries for four important molecular adsorption modes of a cysteine molecule on the FeS$_2$(100) surface. Carbon, hydrogen, oxygen, nitrogen, sulfur, and iron atoms are represented by light gray, white, red, blue, yellow, and dark grey, respectively.

To begin with, we investigated the molecular adsorption (physisorption) of a cysteine molecule on the clean pyrite(100) surface. We obtained ten optimized geometries for the molecular adsorption and found that molecular adsorption was possible (see Fig. S1 and Table S1). Here, we report four energetically favorable adsorption modes (Fig. 2 and Table 1). A first one is the thiol headgroup adsorption (configuration **H1**). The S$_{Cys}$–Fe distance (where S$_{Cys}$ denotes the S atom of the thiol headgroup) was 2.40 Å. The adsorption energy was −7.0 kcal/mol. A second one is an adsorption mode in which the thiol headgroup and the oxygen atom of the carboxyl group are simultaneously involved (configuration **HC1**). In **HC1**, the S$_{Cys}$–Fe interaction appeared somewhat weak (the S$_{Cys}$–Fe distance was 2.67 Å), resulting in the adsorption energy of −7.5 kcal/mol. A third one (configuration **HC2**) is similar to **HC1**; in this case, however, the S$_{Cys}$–Fe distance was shorter (2.51 Å) than that in **HC1**, which indicates a stronger S$_{Cys}$–Fe interaction. As a result of two interacting sites, the cysteine molecule was more strongly adsorbed on the (100) surface: the adsorption energy was −10.5 kcal/mol. A fourth one is an adsorption mode in which the two oxygen atoms of the COO$^-$ group of the zwitterion form interact with the two Fe atoms (configuration **Z**). The distances between the Fe and the oxygen atoms were 2.13 and 2.18 Å. The adsorption energy was −10.6 kcal/mol, which was energetically comparable with that of **HC2**. This may be consistent with a recent experimental study using X-ray photoemission spectroscopy under UHV conditions, where signals assigned for NH$_2$ and NH$_3^+$ groups are simultaneously observed [47] (however, we note that we did not investigate anionic or cationic forms in the present work).

### Table 2

Adsorption energies and structural parameters of thiol headgroup adsorption on the clean and sulfur-deficient $FeS_2(100)$ surfaces. $S_{Cys}$ denotes the sulfur atom of the cysteine molecule and $d(Fe-S)_{ax}$ is the axial Fe–S bond distance on the surface.

| Configuration | $\Delta E_{ads}$ (kcal/mol) | $S_{Cys}$–Fe (Å) | $d(Fe-S)_{ax}$ (Å) |
|---|---|---|---|
| Molecular adsorption | | | |
| **H1** (Clean) | −7.0 | 2.40 | 2.26 |
| **H2** (Sulfur vacancy) | −11.4 | 2.38 | 2.24 |
| Dissociative adsorption | | | |
| **HD1** (Clean) | +13.0 | 2.36 | 2.29 |
| **HD2** (Sulfur vacancy) | −14.1 | 2.15 | 2.25 |

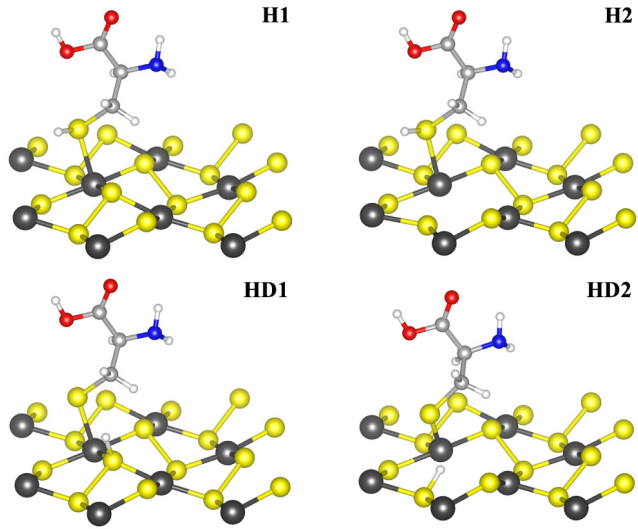

**Fig. 3.** Optimized geometries for important thiol-headgroup adsorption modes on the clean (left) and sulfur-deficient (right) $FeS_2(100)$ surfaces. Upper and lower panels represent physisorption and chemisorption, respectively. Carbon, hydrogen, oxygen, nitrogen, sulfur, and iron atoms are represented by light gray, white, red, blue, yellow, and dark grey, respectively. Only the three uppermost atomic layers of the slab are shown for clarity.

Next, we investigated roles of sulfur vacancies on cysteine adsorption because sulfur vacancies can be found on the surfaces of natural pyrite [1]. In particular, we focused on the headgroup adsorption, which has been one of the important subjects on the cysteine–metal interactions [28, 29, 32, 38]. First, we looked at the case of physisorption (Table 2 and upper panels in Fig. 3). The molecular adsorption energy on the sulfur-vacancy surface (configuration **H2**) was −11.4 kcal/mol, which was found to be more stable by 4.4 kcal/mol than that on the clean surface (**H1**) (Table 2). The $S_{Cys}$–Fe distance was 2.38 Å, which was slightly (0.02 Å) shorter than that in the case of the ideal surface (2.40 Å). The S–H bond length (1.35 Å) on the sulfur-vacancy surface remained almost unchanged compared with that on the ideal surface. To further clarify the roles of surface vacancies, we removed another S atom from the sulfur-vacancy surface and obtained the optimized geometry of cysteine physisorption (not shown): the adsorption energy was −19.6 kcal/mol, which was more stable by 8.2 kcal/mol than that in the case of one-sulfur-vacancy surface. Our results suggest that sulfur-vacancy surfaces tend to stabilize cysteine molecular adsorption.

We also investigated chemisorption (dissociative adsorption) on both the clean and sulfur-vacancy surfaces (Table 2 and lower panels in Fig. 3). We found that the dissociative adsorption of the thiol headgroup on the ideal surface (configuration **HD1**) was energetically unfavorable, which is indicated by the calculated adsorption energy of $+13.0$ kcal/mol. On the contrary, the dissociative adsorption on the sulfur-vacancy surface (configuration **HD2**) appeared energetically favorable; in fact, the adsorption energy was $-14.1$ kcal/mol, which was more stable by 2.7 kcal/mol than that for the physical adsorption (**H2**). The $S_{surf}$–H bond (where $S_{surf}$ denotes S atom of the pyrite surface) was 1.37 Å, which indicates that the $S_{surf}$–H bond appeared to be slightly weaker than that of an isolated cysteine molecule. The $S_{Cys}$–Fe distance was 2.15 Å, being 0.23 Å shorter than that in the case of the molecular adsorption (Table 2), showing a stronger $S_{Cys}$–Fe interaction. We actually obtained ten optimized geometries for the molecular adsorption of a cysteine molecule on the sulfur-deficient surface (see Fig. S2 and Table S2); and we note that the adsorption of the zwitterion form on the sulfur-vacancy surface (configuration **ZV** in Fig. S2, which is similar to **Z**) was more energetically favored by 2.2 kcal/mol than that for **HD2**. Nevertheless, the results suggest that, in the case of the thiol headgroup adsorption on the sulfur-deficient surface, chemisorption is energetically favored than physisorption. This tendency agrees with the previous studies, in which dissociative adsorption on the sulfur-vacancy surface is energetically favorable for a water molecule and glycine [51].

Having investigated cysteine adsorption, we now address the adsorption of its oxidized dimer, cystine, which is composed of two cysteine molecules linked together via the S–S bond. Because of inherent dihedral rotation in a cystine molecule, numerous adsorption modes of cystine on pyrite surface may be possible; however, a first important question appears to be whether cystine adsorption takes the form of thiolate or disulfide [28, 29, 35, 36]. Previous studies on cysteine adsorption on surfaces of noble metals suggest that they tend to form S(thiolate)–metal bonds. For instance, in the case of Au(111), cystine is likely to dissociate, leading to the formation of thiolate S–Au bonds [28, 29]; also, the thiolate–gold bond on the Au(110) surface is energetically possible [32]. In metallic silver surfaces, the formation of the S–Ag bond is possible [37, 38]; on the other hand, in silver colloidal solution, the formation of cystine molecules by the S–S bond may be possible [35, 36]. In the case of adsorption of cysteine on Cu, there is some dimerization of cysteine molecules into cystine [39].

To shed light into this question, we investigated both the physisorption and chemisorption of a cystine molecule on the clean and sulfur-deficient surfaces. Considering that there can be various possible rotational conformers, we first investigated 15 molecular adsorption modes on the clean surface (Fig. S3). However, a comprehensive investigation on adsorption modes of all possible rotational conformers was still prohibitive and thus beyond the scope of the present work. Nonetheless, we found that, in our conformers examined, the disulfide bridge (the S–S bond) of a cystine molecule tended to keep away from the pyrite surface owing to the preferential interactions between the surface and the carboxyl group(s) (Fig. S3).

Table 3

Adsorption energies and structural parameters of cystine adsorption on the clean and defective FeS$_2$(100) surfaces. S$_{cystine}$ denotes the sulfur atom of the cystine molecule and O$_C$ the oxygen atom of the carboxyl group.

| Configuration | Clean | | | Defective | | |
|---|---|---|---|---|---|---|
| | $\Delta E_{ads}$ (kcal/mol) | S$_{cystine}$–Fe (Å) | O$_C$–Fe (Å) | $\Delta E_{ads}$ (kcal/mol) | S$_{cystine}$–Fe (Å) | O$_C$–Fe (Å) |
| P1 | −10.0 | 2.95$^a$ | 2.15 | −6.7 | 2.99$^a$ | 2.11 |
| P2 | −8.5 | 3.57$^a$ | 2.30 | −5.3 | 3.63$^a$ | 2.24 |
| P3 | −7.5 | 3.13$^a$ | – | −5.3 | 2.96$^a$ | – |
| C1 | +2.9 | 2.30$^b$ | 2.02 | −23.0 | 2.21$^b$ | 2.01 |
| C2 | −12.4 | 2.28$^b$ | 2.03$^b$ | −33.0 | 2.22$^b$ | 2.01$^b$ |
| C3 | −15.6 | 2.31$^b$ | – | −23.4 | 2.19$^b$ | – |

$^a$Nearest S$_{cystine}$–Fe distance
$^b$Average value of the two distances

To directly compare the differences between physisorption and chemisorption, here we take six adsorption configurations into account (Table 3). On the defective surface, the molecular adsorption energies were −6.7, −5.3, and −5.3 kcal/mol for configurations **P1**, **P2**, and **P3** (Fig. 4), respectively; on the other hand, the dissociative adsorption energies were −23.0, −33.0, and −23.4 kcal/mol for configurations **C1**, **C2**, and **C3** (Fig. 5), respectively. We note that on the clean surface the dissociative adsorption was still energetically favored for **C2** and **C3** compared with their molecular adsorption counterparts (**P2** and **P3**) (Table 3). In addition, we found that in the case of chemisorption a cystine molecule was more strongly adsorbed on the defective surface in comparison with the clean surface; actually, the S$_{cystine}$–Fe distances became 0.09, 0.06, and 0.12 Å shorter and the adsorption energies were stabilized by 25.9, 20.6, and 7.8 kcal/mol for **C1**, and **C2**, and **C3**, respectively (Table 3). This is essentially in line with the case of cysteine adsorption on the sulfur-vacancy surface, which was discussed earlier. Our results suggest that under vacuum conditions the thiolate S–Fe bond is energetically favorable on the sulfur-deficient FeS$_2$(100) surface.

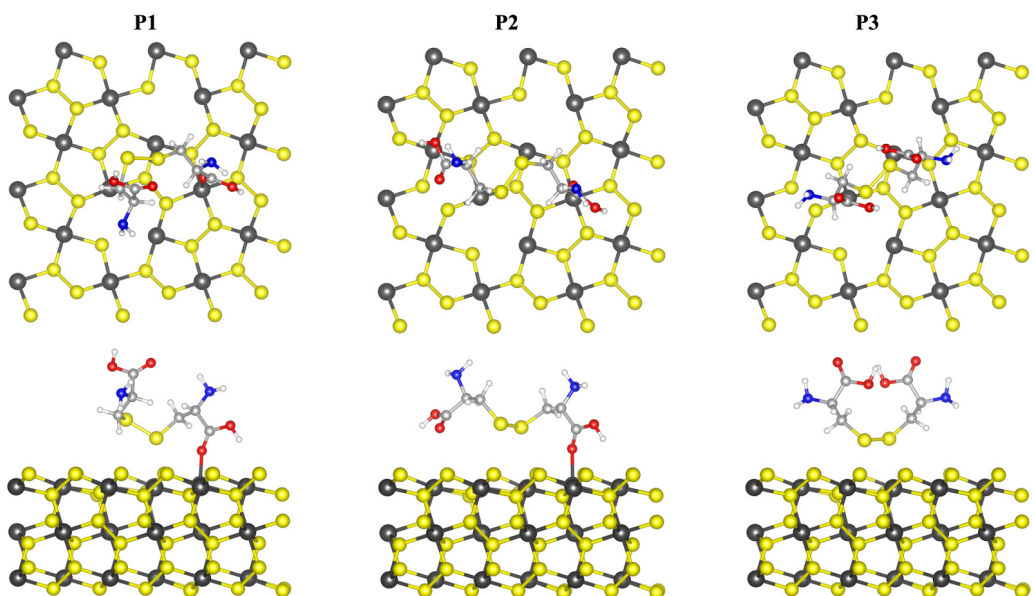

**Fig. 4.** Optimized geometries for three molecular adsorption modes of a cystine molecule on the defective FeS$_2$ (100) surface. Carbon, hydrogen, oxygen, nitrogen, sulfur, and iron atoms are represented by light gray, white, red, blue, yellow, and dark grey, respectively. In the top view (upper panels), only the three uppermost atomic layers of the slab are shown for clarity.

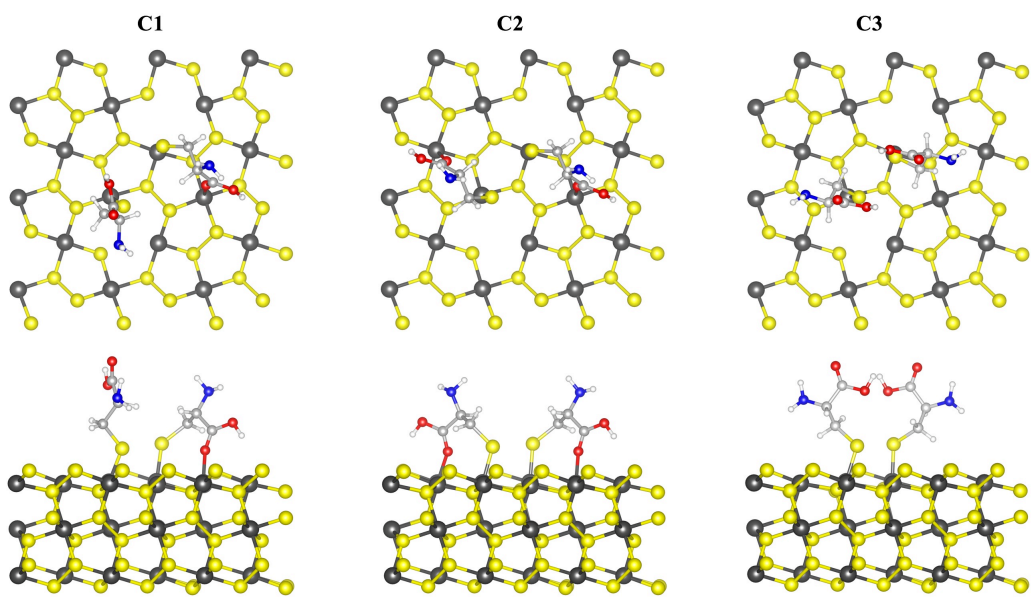

**Fig. 5.** Optimized geometries for three dissociative adsorption modes of a cystine molecule on the defective FeS$_2$ (100) surface. Carbon, hydrogen, oxygen, nitrogen, sulfur, and iron atoms are represented by light gray, white, red, blue, yellow, and dark grey, respectively. In the top view (upper panels), only the three uppermost atomic layers of the slab are shown for clarity.

To understand this, we also investigated the adsorption of dimethyl disulfide (DMDS) (CH$_3$SSCH$_3$) on the clean (100) face of pyrite (Fig. S4 and Table S3). This model system can be viewed as a prototype for cystine adsorption on the FeS$_2$(100) surface. The S$_{DMDS}$–Fe distances were 3.15 and 2.28 Å for molecular and dissociative adsorption, respectively. Consistently, the molecular adsorption energy was –8.1 kcal/mol, whereas the dissociative adsorption energy was −17.5 kcal/mol. This suggests that, while the molecular adsorption of DMDS is still energetically possible, the dissociative adsorption appears to be more favorable. From a geometrical point of view, a possible reason is that the nearest Fe–Fe distance on the (100) face (3.74 Å) is considerably larger than the S–S bond length of an isolated cystine (2.06 Å) (see Fig. S4); consequently, the S–S bond was significantly lengthened from 2.06 to 2.39 Å (a 16% increase). The results support that the thiolate S–Fe bond is energetically favored in the gas phase. To understand this tendency from a different point of view, we also computationally investigated a possible oxidation reaction of two cysteine molecules on the defective surface under vacuum conditions and found that dimerization required at least 31–43 kcal/mol (Figs. S5–S7).

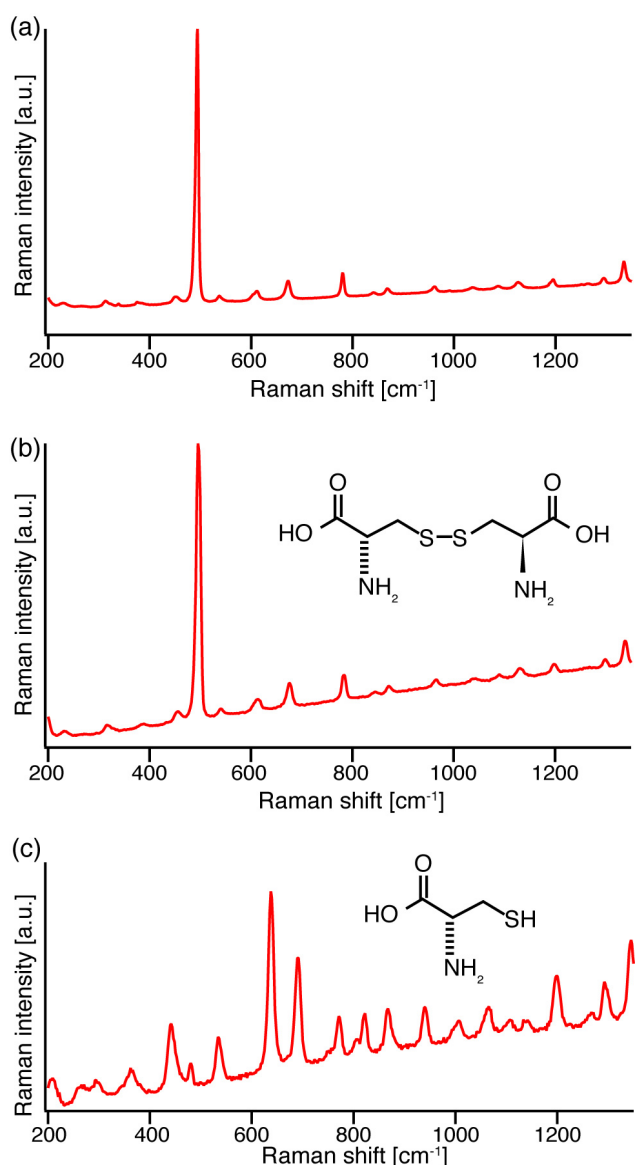

**Fig. 6.** (a) Raman spectrum of cysteine-adsorbed pyrite surface, indicating dimerization on the pyrite surface; (b) Raman spectrum of neat cystine; (c) Raman spectrum of neat cysteine. Note that the Raman spectrum in (a) was measured on the pyrite surface after a natural cubic crystal of pyrite had been dipped into cysteine aqueous solution for 10 hours.

In our DFT calculations, only the adsorption of a single molecule on the pyrite surface was studied; for that reason, our computational results (low coverage in the gas phase) should not be directly compared with experiments on the cysteine adsorption in solutions and/or at high coverage, where self-assembly at pyrite surfaces could occur. With this in mind, we carried out Raman spectroscopy measurements on the cysteine adsorption on the FeS$_2$ surface in aqueous solution. The Raman spectrum was measured on the pyrite surface after a natural cubic crystal of pyrite had been dipped into a 0.1 mM aqueous solution of cysteine for 10 hours. We experimentally observed the formation of cystine molecules through the cysteine adsorption on the pyrite surface in aqueous solution: the strongest Raman band observed at 495 cm$^{-1}$ is attributed to an S–S stretching mode, indicating the disulfide bond of cystine. All the other Raman bands in Fig. 6(a) match nicely with those of neat cystine (Fig. 6(b)) and are clearly differentiated from those of neat cysteine (Fig. 6(c)). Our combined computational and experimental results are in line with the cases of the cysteine–Ag interactions, in which the formation of the S–Ag bond is possible in metallic silver surfaces [37, 38] whereas the formation of cystine molecules through the S–S bond may be possible in silver colloidal solution [35, 36]. Future computational work should address the cysteine and cystine adsorption in aqueous solution and/or at high coverage.

The results presented here support the hypothesis that pyrite could have played an important role as a concentrator of amino acids under anoxic conditions in prebiotic chemistry [14–23]. In particular, a variety of possible adsorption modes of amino acids (as was demonstrated for cysteine adsorption in this study) could facilitate different polymerization of organic molecules, possibly leading to complex biological molecules. Further, our combined computational and experimental results might have implications for the origins of iron–sulfur proteins and redox chemistry on the early Earth. Although we could not find a direct link between cysteine adsorption on pyrite surfaces and iron–sulfur clusters (where the Fe atom is tetrahedrally coordinated, leading to different spin states [24–26, 56–58]), it may be hypothesized that cysteine adsorption on pyrite surfaces could have played a significant role in the emergence and evolution of iron–sulfur proteins. We hope that further experimental and computational studies provide deeper insights into this open question.

## 4. Conclusions

In the present work, we have computationally studied the adsorption of L-cysteine and its oxidized dimer, L-cystine, on the clean and defective FeS$_2$(100) surfaces in the gas phase. Our calculations suggest that sulfur-deficient surfaces play an important role in the adsorption of both cysteine and cystine. In particular, we have found that, in the thiol headgroup adsorption on the sulfur-vacancy site, chemisorption is more energetically favored compared with physisorption. Our calculations also indicate that, in the cystine adsorption on the defective surface under vacuum conditions, the formation of the S–Fe bond is energetically favorable compared with molecular adsorption. This tendency was supported by our calculations on DMDS adsorption on the ideal surface. In addition, we have carried out Raman spectroscopy measurements on the cysteine adsorption on the FeS$_2$ surface. We have experimentally observed the formation of cystine molecules on the pyrite surface in aqueous solution. Future computational work should address the cysteine and cystine adsorption in aqueous solution and/or at high coverage. Our results might have implications for chemical evolution at mineral surfaces at the early Earth and the origin of iron–sulfur proteins.

Supplementary materials

Supplementary material associated with this article can be found, in the online version.

## Acknowledgments

We thank Yousoo Kim (RIKEN) for helpful discussions. This work was partially supported by "Exploratory Challenge on Post-K computer" (Frontiers of Basic Science: Challenging the Limits) by the Ministry of Education, Culture, Sports, Science, and Technology (MEXT), Japan. T.E. was supported in part by Grant-in-Aid for Scientific Research on Innovative Areas, "Hadean Bioscience" (Grant No. 26106006) and Grant-in-Aid for Advanced Utilization of High Performance General-Purpose Computer (2017). T.Y. and M.H. were supported in part by JSPS-KAKENHI Grant-in-Aid for Scientific Research on Innovative Areas "Hadean Bioscience", Grant Number JP26106003. The calculations were performed on HOKUSAI GreatWave at Advanced Center for Computing and Communication, RIKEN.

Supplementary Material for "Cysteine and cystine adsorption on FeS$_2$(100)"

Teppei Suzuki, Taka-aki Yano, Masahiko Hara, and Toshikazu Ebisuzaki

## S1. Cysteine adsorption on the clean FeS$_2$(100) surface

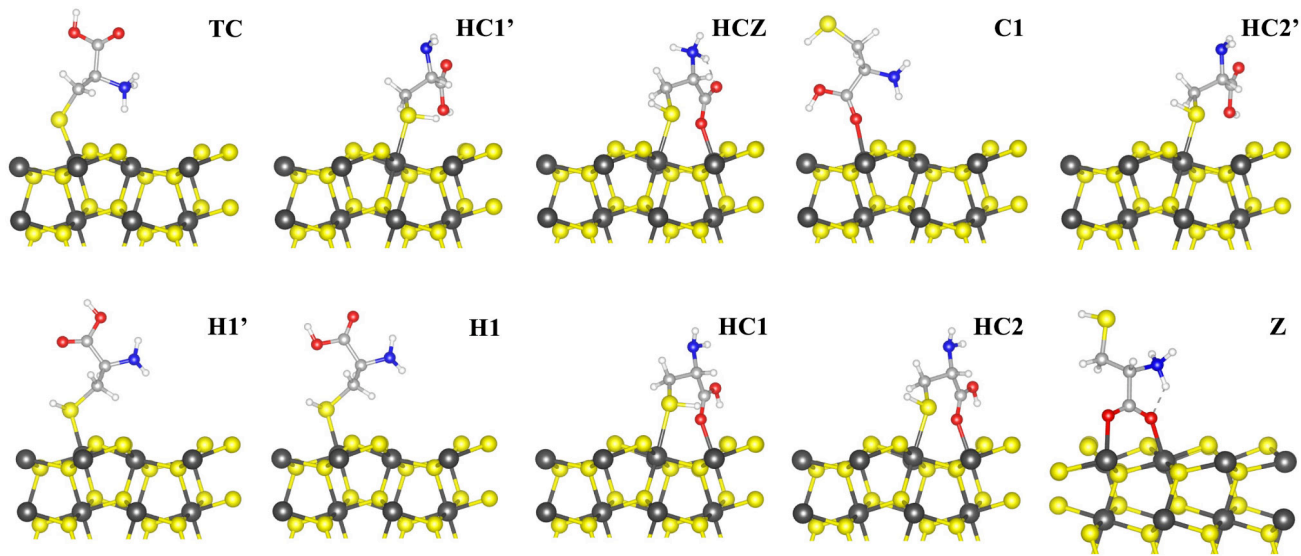

**Fig. S1.** Optimized geometries for 10 selected molecular adsorption modes of a cysteine molecule on the clean FeS$_2$(100) surface. Carbon, hydrogen, sulfur, and iron atoms are represented by light gray, white, yellow, and dark grey, respectively. Only the upper layers of the slab are shown

**Table S1:** Adsorption energies and structural parameters of cysteine adsorption on the clean FeS$_2$(100) surface. S$_{Cys}$ denotes the sulfur atom of the cysteine molecule and O$_C$ the oxygen atom of the carboxyl group.

| Configuration | $\Delta E_{ads}$ (kcal/mol) | S$_{Cys}$–Fe (Å) | O$_C$–Fe (Å) |
|---|---|---|---|
| TC   | −1.1  | 2.36 | –    |
| HC1' | −4.0  | 2.40 | 2.93 |
| HCZ  | −4.1  | 2.42 | 2.04 |
| C1   | −5.0  | –    | 2.15 |
| HC2' | −6.0  | 2.39 | 2.77 |
| H1'  | −6.1  | 2.41 | –    |
| H1   | −7.0  | 2.40 | –    |
| HC1  | −7.5  | 2.67 | 2.02 |
| HC2  | −10.5 | 2.51 | 2.03 |
| Z    | −10.6 | –    | 2.13, 2.18 |

## S2. Cysteine adsorption on the sulfur-vacancy FeS$_2$(100) surface

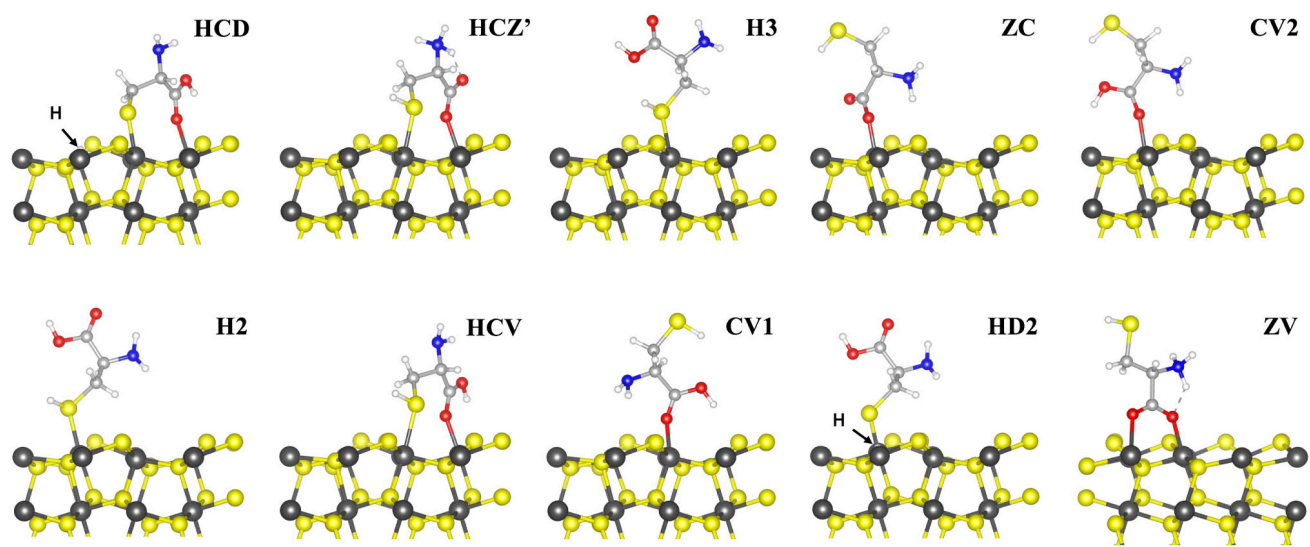

**Fig. S2.** Optimized geometries for 10 selected molecular adsorption modes of a cysteine molecule on the sulfur-vacancy FeS$_2$(100) surface. Carbon, hydrogen, sulfur, and iron atoms are represented by light gray, white, yellow, and dark grey, respectively. Only the upper layers of the slab are shown.

**Table S2**: Adsorption energies and structural parameters of cysteine adsorption on the sulfur-vacancy FeS$_2$(100) surface. S$_{Cys}$ denotes the sulfur atom of the cysteine molecule and O$_C$ the oxygen atom of the carboxyl group.

| Configuration | $\Delta E_{\mathrm{ads}}$ (kcal/mol) | S$_{Cys}$–Fe (Å) | O$_C$–Fe (Å) |
|---|---|---|---|
| HCD   | $-5.5$  | 2.21 | 2.04 |
| HCZ'  | $-5.6$  | 2.46 | 2.03 |
| H3    | $-6.5$  | 2.39 | –    |
| ZC    | $-9.7$  | –    | 1.97 |
| CV2   | $-10.5$ | –    | 2.11 |
| H2    | $-11.4$ | 2.38 | –    |
| HCV   | $-11.6$ | 2.56 | 2.03 |
| CV1   | $-13.7$ | –    | 1.97 |
| HD2   | $-14.1$ | 2.15 | –    |
| ZV    | $-16.3$ | –    | 2.07, 2.17 |

## S3. Molecular adsorption of a cystine molecule on the clean FeS$_2$(100) surface

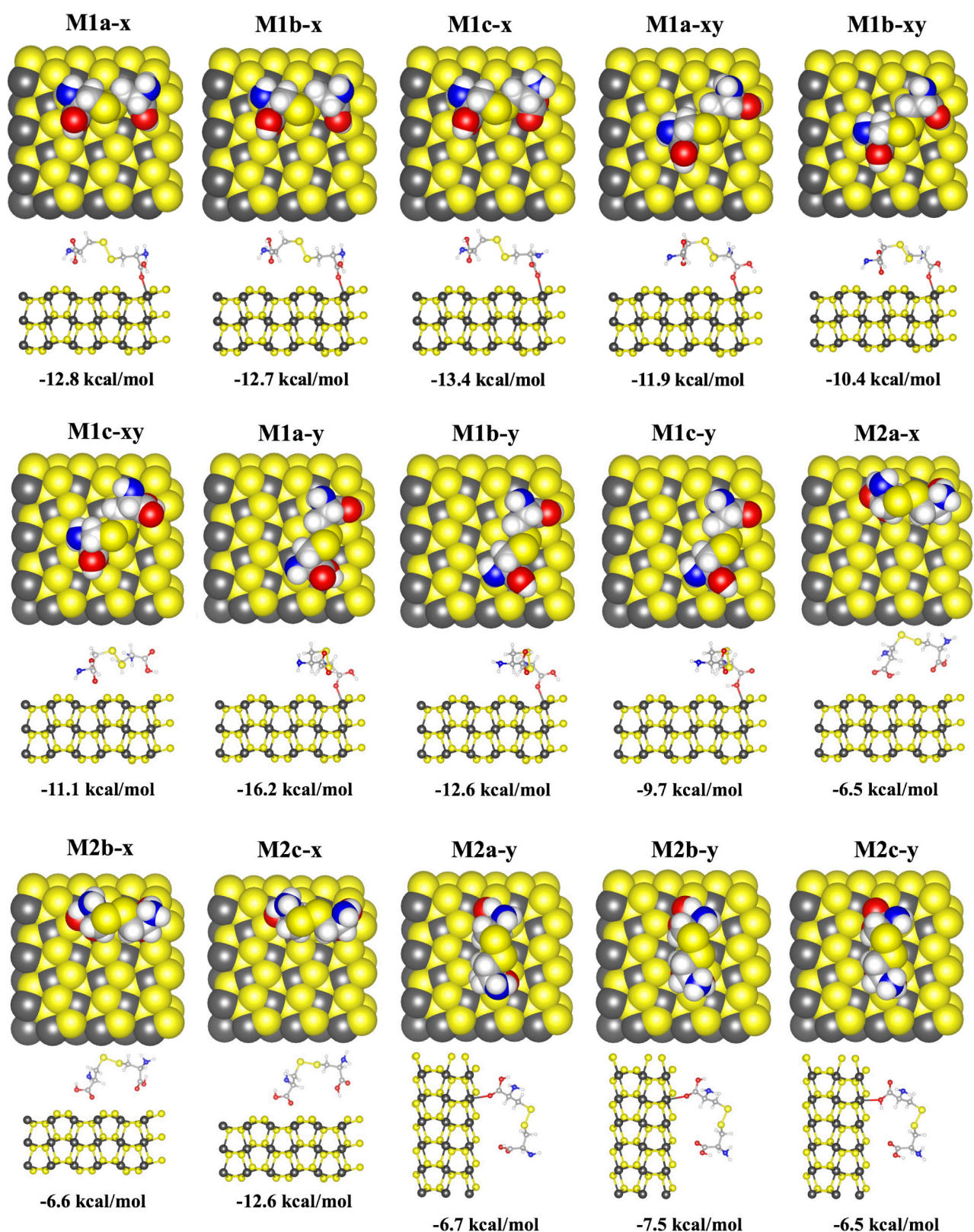

**Fig. S3.** Optimized geometries for selected molecular adsorption modes of a cystine molecule on the clean FeS$_2$(100) surface. Carbon, hydrogen, oxygen, nitrogen, sulfur, and iron atoms are represented by light gray, white, red, blue, yellow, and dark grey, respectively.

## S4. Adsorption of a dimethyl disulfide (DMDS) on the clean FeS$_2$(100) surface

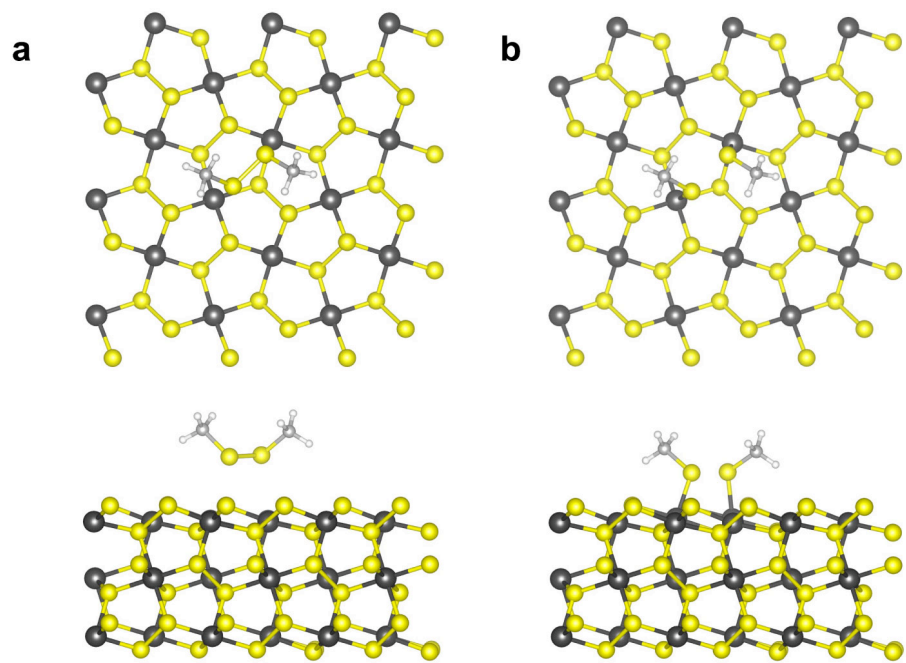

**Fig. S4.** Optimized geometries for the adsorption of a DMDS molecule on the clean FeS$_2$(100) surface: Molecular adsorption (a) and dissociative adsorption (b). Carbon, hydrogen, sulfur, and iron atoms are represented by light gray, white, yellow, and dark grey, respectively. In the top view (upper panels), only the three uppermost atomic layers of the slab are shown for clarity.

**Table S3:** Adsorption energies and structural parameters for DMDS adsorption on the clean FeS$_2$(100) surface. The distance $d$(S–S)$_{\text{DMDS}}$ is the S–S distance of the adsorbed DMDS molecule and the changes from that of an isolated molecule are given in parentheses.

| Adsorption | $\Delta E_{\text{ads}}$ (kcal/mol) | S$_{\text{DMDS}}$–Fe(Å) | $d$(S–S)$_{\text{DMDS}}$ (Å) |
|---|---|---|---|
| Molecular adsorption | −8.1 | 3.15 | 2.09 (+1.5%) |
| Dissociative adsorption | −17.5 | 2.28 | 2.39 (+16%) |

## S5. Adsorption and oxidation of two cysteine molecules on the defective FeS$_2$(100) surface

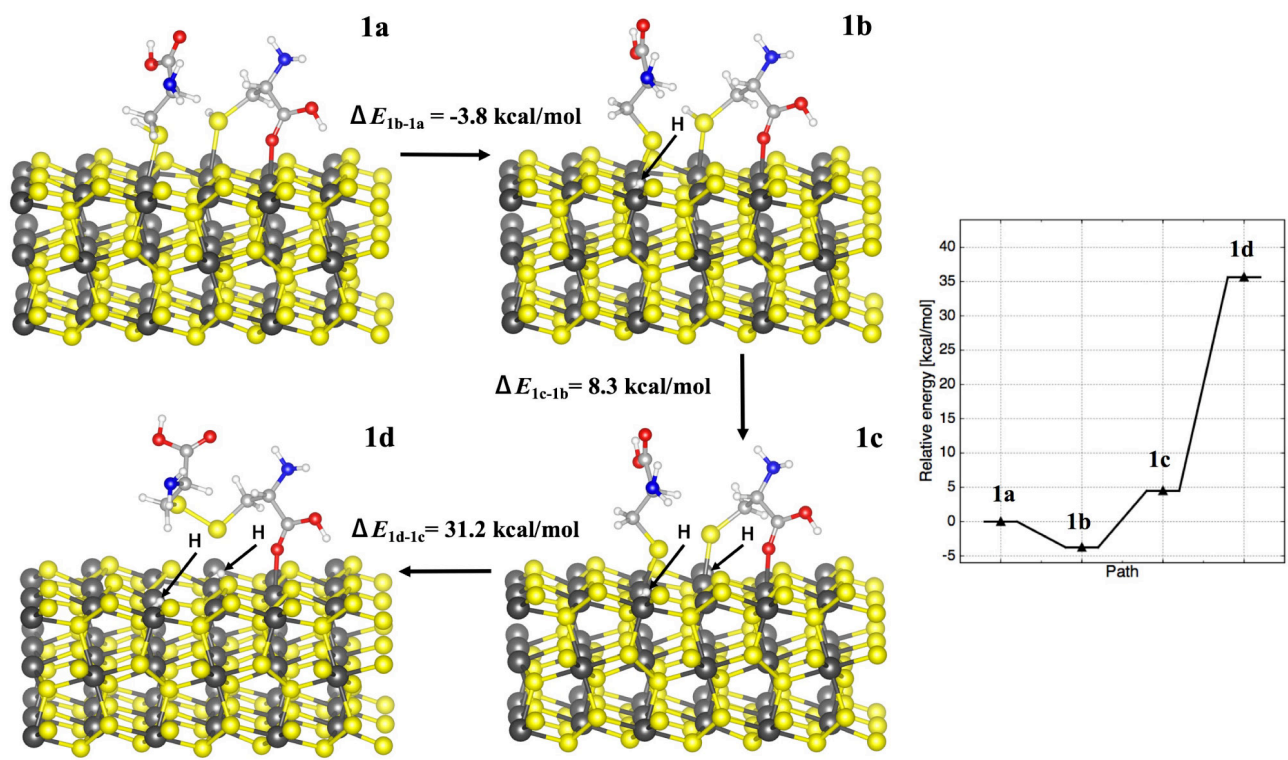

**Fig. S5.** Conjectured oxidization mechanism of two cysteine molecules (in configuration **1a**) on the defective FeS$_2$(100) surface. Carbon, hydrogen, oxygen, nitrogen, sulfur, and iron atoms are represented by light gray, white, red, blue, yellow, and dark grey, respectively. In the beginning, two neutral cysteine molecules are absorbed on the pyrite surface (**1a**). Then, an S–H bond of a cysteine molecule is dissociated (configuration **1b**); and the other S–H bond is subsequently dissociated (configuration **1c**). Finally, a cystine molecule is formed and two H atoms are left on the surface (configuration **1c**), requiring the relative energy of 31.2 kcal/mol.

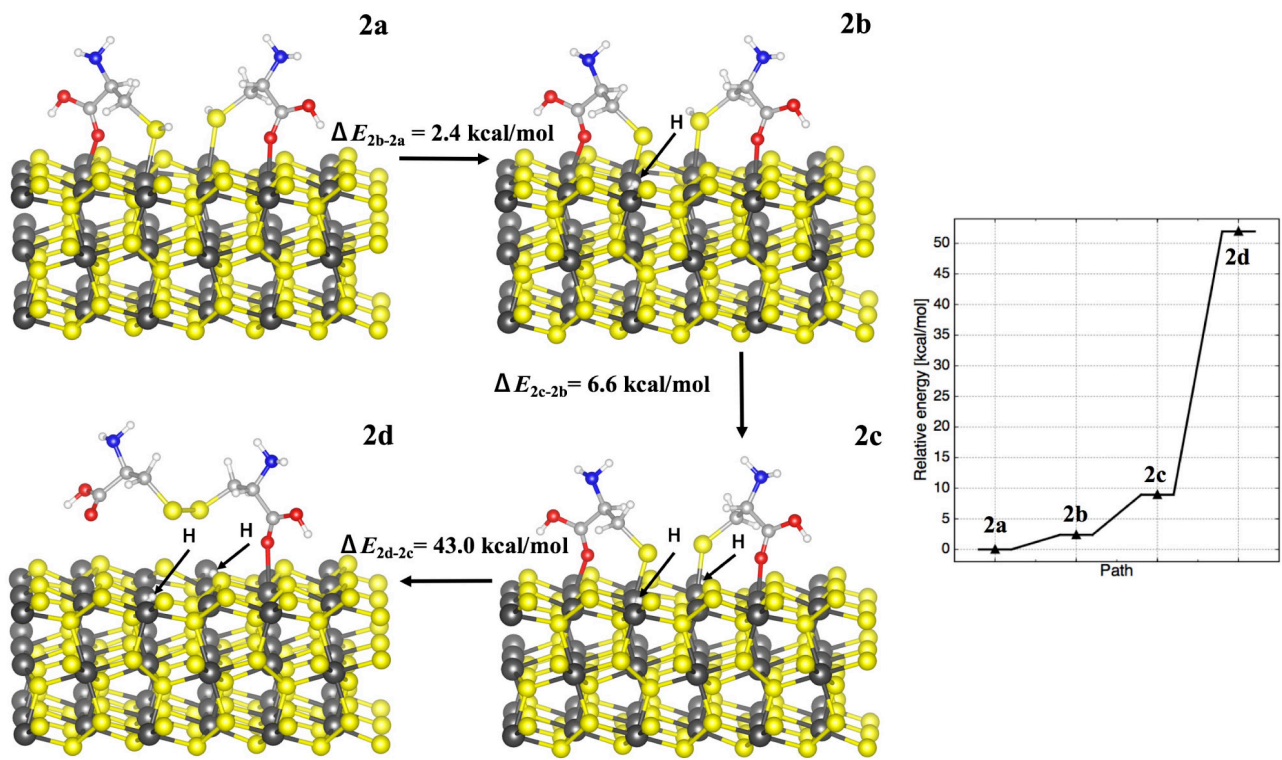

**Fig. S6.** Conjectured oxidization mechanism of two cysteine molecules (in configuration **2a**) on the defective FeS$_2$(100) surface. Carbon, hydrogen, oxygen, nitrogen, sulfur, and iron atoms are represented by light gray, white, red, blue, yellow, and dark grey, respectively. In the beginning, two neutral cysteine molecules are absorbed on the pyrite surface (**2a**). Then, an S–H bond of a cysteine molecule is dissociated (configuration **2b**); and the other S–H bond is subsequently dissociated (configuration **2c**). Finally, a cystine molecule is formed and two H atoms are left on the surface (configuration **2c**), requiring the relative energy of 43.0 kcal/mol.

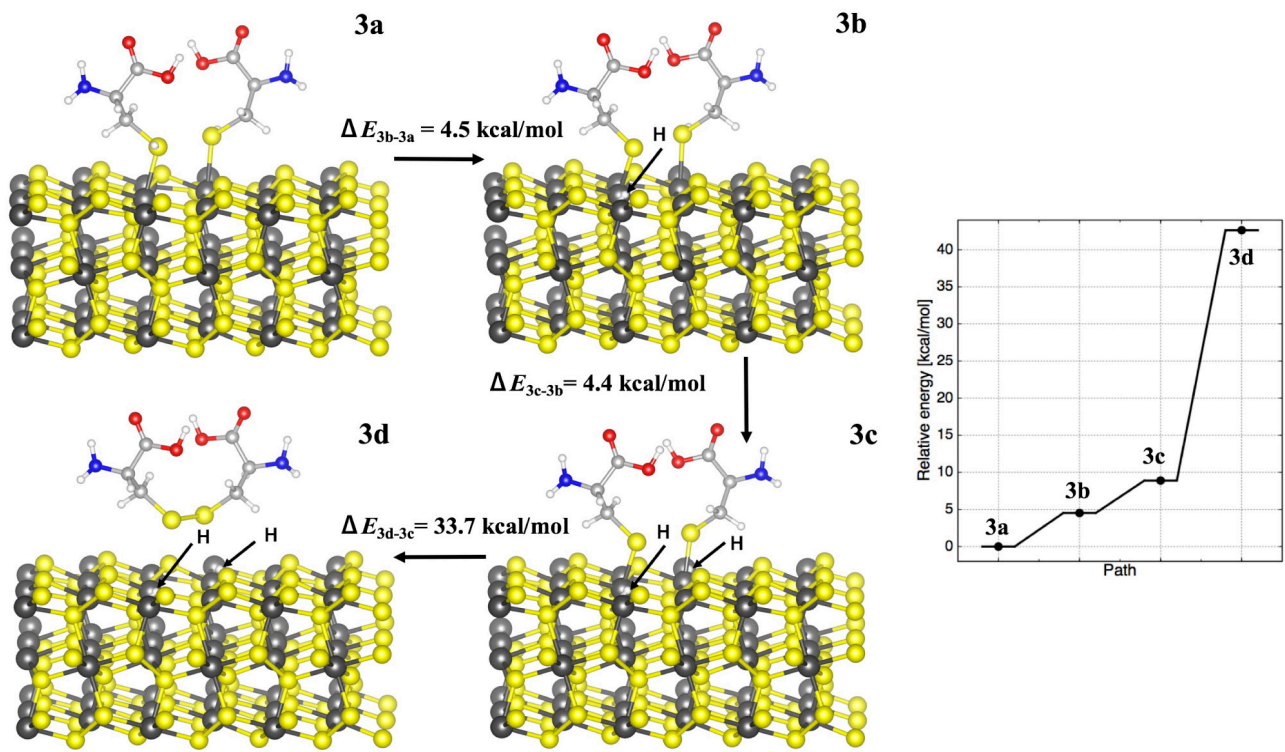

**Fig. S7.** Conjectured oxidization mechanism of two cysteine molecules (in configuration **3a**) on the defective FeS$_2$(100) surface. Carbon, hydrogen, oxygen, nitrogen, sulfur, and iron atoms are represented by light gray, white, red, blue, yellow, and dark grey, respectively. In the beginning, two neutral cysteine molecules are absorbed on the pyrite surface (**3a**). Then, an S–H bond of a cysteine molecule is dissociated (configuration **3b**); and the other S–H bond is subsequently dissociated (configuration **3c**). Finally, a cystine molecule is formed and two H atoms are left on the surface (configuration **3c**), requiring the relative energy of 33.7 kcal/mol.